\documentclass[%
	aps,
	prd,
	reprint,
	superscriptaddress,
	floatfix,
	nofootinbib,
	noamsmath,
	noamssymb
]
{revtex4-2}

\usepackage[T1]{fontenc}
\usepackage[utf8]{inputenc}
\usepackage{graphicx}
\usepackage{xcolor}
\usepackage{mathtools}
\usepackage{amssymb}
\usepackage{lmodern}
\usepackage{bm}
\usepackage{siunitx}
\usepackage{slashed}
\usepackage[final]{microtype}
\usepackage{hyperref}

\newcommand{\mytitle}{Thermodynamics from the quark condensate}

\newcommand{\JLU}{%
	Institut f\"{u}r Theoretische Physik, %
	Justus-Liebig-Universit\"{a}t Gie\ss{}en, %
	35392 Gie\ss{}en, %
	Germany%
}

\newcommand{\HFHF}{%
	Helmholtz Forschungsakademie Hessen f\"{u}r FAIR (HFHF), %
	GSI Helmholtzzentrum\\f\"{u}r Schwerionenforschung, %
	Campus Gie\ss{}en, %
	35392 Gie\ss{}en, %
	Germany%
}

\newcommand{\DWD}{%
	Deutscher Wetterdienst, %
	Frankfurter Str.~135, %
	63067 Offenbach am Main, %
	Germany%
}

\hypersetup{%
	pdftitle = {\mytitle},
	pdfauthor = {Philipp Isserstedt, Christian S. Fischer, Thorsten Steinert},
	bookmarksopen = true,
	colorlinks = true,
	linkcolor = red!50!black,
	citecolor = blue!50!black,
	urlcolor = blue!50!black
}

\graphicspath{{./Graphics/}}

\DeclarePairedDelimiter{\abs}{\lvert}{\rvert}
\DeclarePairedDelimiter{\expval}{\langle}{\rangle}

\newcommand{\bbZ}{\mathbb{Z}}
\newcommand{\calB}{\mathcal{B}}
\newcommand{\calL}{\mathcal{L}}
\newcommand{\calZ}{\mathcal{Z}}
\newcommand{\+}{\hspace*{.08335em}}
\newcommand{\beq}{\begin{equation}}
\newcommand{\eeq}{\end{equation}}
\newcommand{\dd}{\mathrm{d}}
\newcommand{\ii}{\mathrm{i}}
\newcommand{\ee}{\operatorname{e}}
\newcommand{\Tr}{\operatorname{Tr}}
\newcommand{\upu}{\textup{u}}
\newcommand{\upd}{\textup{d}}
\newcommand{\ups}{\textup{s}}
\newcommand{\upB}{\textup{B}}
\newcommand{\qcond}{\expval{\bar{\psi} \psi}}
\newcommand{\ucond}{\qcond_{\upu}}

\newcommand{\scond}{\qcond_{\ups}}
\newcommand{\massu}{m_{\upu}}
\newcommand{\massd}{m_{\upd}}
\newcommand{\masss}{m_{\ups}}
\newcommand{\muu}{\mu_{\upu}}
\newcommand{\mud}{\mu_{\upd}}
\newcommand{\mus}{\mu_{\ups}}
\newcommand{\muc}{\mu_{\textup{c}}}
\newcommand{\mucN}{\muc^{\text{N}}}
\newcommand{\mucfirst}{\muc^{\text{1st}}}
\newcommand{\muB}{\mu_{\upB}}
\newcommand{\Nf}{N_{\textup{f}}}
\newcommand{\Nc}{N_{\textup{c}}}
\newcommand{\Tc}{T_{\textup{c}}}
\newcommand{\SYM}{s_{\textup{YM}}}
\newcommand{\pN}{p_{\textup{N}}}
\newcommand{\pW}{p_{\textup{W}}}
\newcommand{\sN}{s_{\textup{N}}}
\newcommand{\sW}{s_{\textup{W}}}
\newcommand{\nN}{n_{\textup{N}}}
\newcommand{\nW}{n_{\textup{W}}}
\newcommand{\Deltamuz}{\Delta_{\mu = 0}}
\newcommand{\Deltamunz}{\Delta_{\mu \neq 0}}
\newcommand{\TCEP}{T_{\textup{CEP}}}
\newcommand{\muCEP}{\mu_{\textup{CEP}}}
\newcommand{\OmegaNJL}{\Omega_{\textup{NJL}}^{\textup{(mf)}}}
\providecommand*{\wtilde}[1]{\tilde{#1}}
\renewcommand*{\vec}[1]{\bm{#1}}

\DeclareSIUnit{\MeV}{\mega\electronvolt}
\DeclareSIUnit{\GeV}{\giga\electronvolt}

\begin{document}

\sisetup{separate-uncertainty}

\title{\mytitle}

\author{Philipp Isserstedt}
\email{philipp.isserstedt@physik.uni-giessen.de}
\affiliation{\JLU}
\affiliation{\HFHF}

\author{Christian S.~Fischer}
\email{christian.fischer@theo.physik.uni-giessen.de}
\affiliation{\JLU}
\affiliation{\HFHF}

\author{Thorsten Steinert}
\altaffiliation{Present address:~\DWD}
\affiliation{\JLU}

\begin{abstract}
We present a method to compute thermodynamic quantities within functional
continuum frameworks that is independent of the employed truncation. As a
proof of principle, we first apply it to a Nambu-Jona-Lasinio model in
mean-field approximation. Then, we use the method with solutions obtained from
a coupled set of truncated Dyson-Schwinger equations for the quark and gluon
propagators of ($2 + 1$)-flavor quantum chromodynamics in Landau gauge to obtain
the pressure, entropy density, energy density, and interaction measure across
the phase diagram of strong-interaction matter. We also discuss the limitation
of the proposed method.
\end{abstract}

\maketitle

\section{\label{sec:introduction}Introduction}

The thermal properties of strong-interaction matter described by the theory
of quantum chromodynamics (QCD) at nonvanishing temperature and density are
subject to both experimental and theoretical efforts
\cite{BraunMunzinger:2009zz,Fukushima:2010bq}. Future as well as already
operating heavy-ion-collision facilities such as CBM at FAIR/GSI, NICA at
JINR, and RHIC at BNL aim to probe the phase structure of QCD
\cite{Friman:2011zz,NICA:whitepaper,Aggarwal:2010cw,STAR:whitepaper,
Bzdak:2019pkr}. In order to understand and explain the various phases and
thermodynamic properties of QCD, one needs the pressure, entropy density, and
energy density---in short, the equation of state (EoS)---as a function of
temperature and chemical potential. More generally, the knowledge of the EoS of
matter under a given physical environment is important for numerous reasons
ranging from hydrodynamic simulations in the context of heavy-ion collisions to
astrophysical issues like supernovae or neutron stars; see, e.g.,
Refs.~\cite{Lattimer:2015nhk,Oertel:2016bki} for review articles.

On the theoretical side and at vanishing chemical potential, first-principle
lattice-regularized QCD provides the following well-established picture: the
confined hadronic low-temperature phase characterized by dynamical chiral
symmetry breaking is connected through an analytic crossover to the deconfined
high-temperature phase of the quark-gluon plasma with (partially) restored
chiral symmetry \cite{Aoki:2006we,Aoki:2009sc,Borsanyi:2010bp, 
Bazavov:2011nk,Bhattacharya:2014ara}. Furthermore, the full
continuum-extrapolated EoS is available 
\cite{Borsanyi:2013bia,Bazavov:2014pvz}. The situation is fundamentally
different at nonvanishing (real) chemical potential. There, lattice QCD is
hampered by the sign problem. Results for the EoS are
limited to rather small chemical potential and usually obtained via
extrapolation from imaginary chemical potential or using the  Taylor-expansion
technique \cite{Borsanyi:2012cr,Bellwied:2015rza,Bazavov:2017dus,
Bazavov:2018mes}. Thus, other approaches are necessary to complement the
lattice calculations.

Dyson-Schwinger equations (DSEs) and the functional renormalization group
(FRG) constitute complementary functional continuum frameworks, which are
well-suited to study QCD at nonzero temperature and chemical potential. In
recent years, substantial progress were made both within  QCD and using
low-energy effective models; see, e.g., Refs.~\cite{Fischer:2012vc,
Muller:2013pya,Muller:2013tya,Fischer:2014ata,Fischer:2014vxa,
Isserstedt:2019pgx,Gunkel:2019xnh,Gunkel:2020wcl,Gao:2020qsj,Gao:2020fbl,
Herbst:2013ail,Fu:2016tey,Fu:2018qsk,Fu:2019hdw,Otto:2019zjy,Braun:2020ada}
and the comprehensive reviews \cite{Fischer:2018sdj,Dupuis:2020fhh} as well
as references therein.

In this work, we aim to compute thermodynamic quantities within the DSE
approach. To this end, one needs basically the thermodynamic potential.
In contrast to the FRG where solving its flow equation yields directly the
thermodynamic potential, accessing this quantity within the DSE framework is
extremely difficult and limited to simple models \cite{Blaschke:1997bj,
Xu:2015jwa,Gao:2015kea}. Generally speaking, this is due to the fact that the
DSE approach starts with the first derivative of the thermodynamic potential,
and an integration is needed to get hold of the potential itself. 
Unfortunately, this integration is only possible for certain truncations.
It is thus desirable to develop a truncation-independent way to calculate
thermodynamic quantities from DSEs. The Purpose of this work is to introduce
such a method based on a general relation between the quark condensate and
the entropy density.

The remainder of this paper is organized as follows: In Sec.~\ref{sec:method},
we detail the aforementioned method and apply it to a Nambu-Jona-Lasinio (NJL)
model in Sec.~\ref{sec:njl} as a testing ground to gauge the method's
effectiveness. In Sec.~\ref{sec:dse}, we summarize our DSE framework, which
solutions serve as input to obtain the pressure, entropy density, energy
density, and interaction measure in ($2 + 1$)-flavor QCD. In
Sec.~\ref{sec:results}, we present and discuss our results for these quantities
and finally conclude in  Sec.~\ref{sec:summary}.

\section{\label{sec:method}Thermodynamics and quark condensate}

The fundamental quantity for QCD thermodynamics at nonzero temperature $T$
and quark chemical potential $\mu$ is the thermodynamic potential
\beq
	\label{eq:omega}
	\Omega(T, \mu) = -\frac{T}{V} \log \calZ(T, \mu) \, ,
\eeq
where $\calZ$ denotes the grand-canonical partition function. Here, $V$ is the
volume of the system, and we consider only one light flavor first. Thermodynamic
quantities like pressure ($p$), entropy density ($s$), and number
density ($n$) follow from the standard relations
\beq
\begin{gathered}
	\label{eq:p_s_n}
	p(T, \mu)
	=
	- \bigl( \Omega(T, \mu) - \Omega(0, 0) \bigr) \+ ,
	\\[0.5em]
	s(T, \mu)
	=
	-\frac{\partial \+ \Omega(T, \mu)}{\partial T} \, ,
	\quad
	n(T, \mu)
	=
	-\frac{\partial \+ \Omega(T, \mu)}{\partial \mu} \, .
\end{gathered}
\eeq
Furthermore, a Legendre transform of the pressure yields the energy density
\beq
	\label{eq:epsilon}
	\varepsilon(T, \mu)
	=
	T s(T, \mu) + \mu \+ n(T, \mu) - p(T, \mu) \, ,
\eeq
and with that, one defines the interaction measure
\beq
	\label{eq:I}
	I(T, \mu)
	=
	\varepsilon(T, \mu) - 3 \+ p(T, \mu) \, .
\eeq
It is related to the trace of QCD's energy-momentum tensor (hence, also referred
to as the trace anomaly) and measures the deviation of the EoS from the one of
an ideal gas given by $\varepsilon = 3 \+ p$.

In addition to temperature and chemical potential, the current-quark mass $m$
can be seen as an additional variable the thermodynamic potential depends on.
It appears as an external source for the field bilinear $\bar{\psi} \psi$ in
the QCD action, and the quark condensate is obtained via
\beq
	\label{eq:condensate_def}
	\qcond(T, \mu; m)
	=
	\frac{\partial \+ \Omega(T, \mu; m)}{\partial m} \, .
\eeq
In principle, this relation can be inverted to obtain the thermodynamic
potential as an integral of the quark condensate with respect to the
current-quark mass, i.e.,
\beq
	\label{eq:omega_mass_integral_condensate}
	\Omega(T, \mu; m_{2}) - \Omega(T, \mu; m_{1})
	=
	\int_{m_{1}}^{m_{2}} \dd m' \, \qcond(T, \mu; m') \, .
\eeq
Unfortunately, this relation is not suitable for an actual calculation since
the thermodynamic potential and the quark condensate are both divergent. The
divergence is caused by the vacuum contribution contained in $\Omega$ and is
even present in the noninteracting theory \cite{Kapusta:2006pm}. Since the
divergence is independent of temperature and chemical potential, suitable
derivatives of the potential and condensate are expected to be finite.%
\footnote{Even though the divergent contribution is independent of temperature
and chemical potential, it depends on the current-quark mass. Thus,
$\qcond = \partial \+ \Omega / \partial m$ is divergent, too.}
Therefore, differentiating Eq.~\eqref{eq:omega_mass_integral_condensate} with
respect to $T$ yields the well-defined (divergence-free) equation
\beq
	\label{eq:entropy_density}
	s(T, \mu; m_{2}) - s(T, \mu; m_{1})
	=
	-\int_{m_{1}}^{m_{2}} \dd m' \,
	\frac{\partial \qcond}{\partial T}(T, \mu; m') \, .
\eeq

In order to use this relation in practical calculations, we have to specify the
integral boundaries. We set the lower one to the physical current-quark mass,
$m_{1} = m$, and send the upper one to infinity, $m_{2} \to \infty$. An
infinitely heavy quark freezes out of the system and does not contribute to
thermodynamics. The corresponding entropy density is then simply the one of
pure Yang-Mills theory: $s(T, \mu; m_{2} \to \infty) = \SYM(T)$. Thus, our
final expression for the entropy density reads
\beq
	\label{eq:entropy_density_final}
	s(T, \mu; m)
	=
	\SYM(T)
	+
	\int_{m}^{\infty} \dd m' \,
	\frac{\partial \qcond}{\partial T}(T, \mu; m') \, .
\eeq
If gluons are no active degrees of freedom, like, e.g., in the NJL model,
the Yang-Mills contribution is set to zero. For QCD, $\SYM$ is
taken from the lattice \cite{Boyd:1996bx,Borsanyi:2012ve}. Note that
Eq.~\eqref{eq:entropy_density_final} implies
$\partial s / \partial m = -\partial \qcond / \partial T$,
which is nothing but the Maxwell-like relation
\beq
	\label{eq:maxwell}
	\frac{\partial^2 \+ \Omega(T, \mu; m)}{\partial m \+ \partial T}
	=
	\frac{\partial^2 \+ \Omega(T, \mu; m)}{\partial T \+ \partial m} \, .
\eeq

Having the entropy density at hand, the pressure at vanishing chemical
potential follows thermodynamically consistent from
\beq
	\label{eq:pressure_zero_mu}
	p(T, 0)
	=
	p(T_{0}, 0)
	+
	\int_{T_{0}}^{T} \dd T\+' \, s(T\+', 0) \, ,
\eeq
and an additional integration over the number density yields the pressure at
nonvanishing chemical potential,
\beq
	\label{eq:pressure_nonzero_mu}
	p(T, \mu)
	=
	p(T_{0}, 0)
	+
	\int_{T_{0}}^{T} \dd T\+' \, s(T\+', 0)
	+
	\int_{0}^{\mu} \dd \mu' \, n(T, \mu') \, .
\eeq
Analogously to $\SYM$, the value $p(T_{0}, 0)$ at a reference temperature
$T_{0}$ is treated as an input parameter and taken from the lattice
\cite{Borsanyi:2013bia,Bazavov:2014pvz}.

\begin{figure}[t]
	\centering%
	\includegraphics{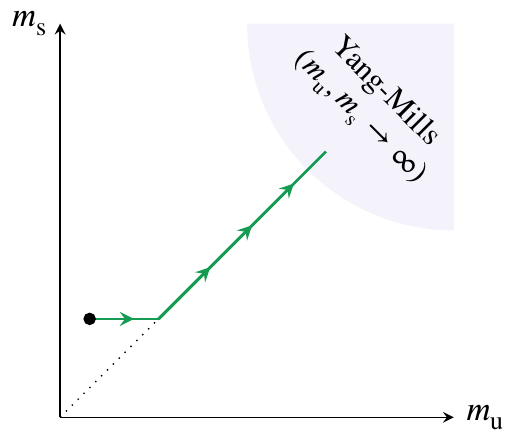}%
	\vspace{-1mm}%
	\caption{\label{fig:Columbia}%
		Sketch of the integration path (green) in the Columbia plot for the
		entropy density in the $(2 + 1)$-flavor case. The black solid circle
		denotes the physical point.
	}
\end{figure}

In the case of $2 + 1$ flavors we have to consider two degenerate light quarks
with $\massu = \massd$, $\muu = \mud$ and a heavier strange quark with a
mass $\masss \gg \massu$. To calculate the entropy density we
integrate within the Columbia plot as follows: first, the up-quark mass from the
physical point to the physical strange-quark mass and then both masses to
infinity (see Fig.~\ref{fig:Columbia}). With $\{ \mu \} = (\muu, \mus)$,
the generalization of Eq.~\eqref{eq:entropy_density_final} reads
\beq
\begin{aligned}
	\label{eq:entropy_density_final_2+1}
	s(T, \{ \mu \})
	&=
	\SYM(T)
	\\[0.25em]
	&\phantom{=\;}
	+
	2 \+ \int_{\massu}^{\masss} \dd m' \,
	\frac{\partial \ucond}{\partial T}(T, \{ \mu \}; m', \masss)
	\\[0.25em]
	&\phantom{=\;}
	+
	2 \+ \int_{\masss}^{\infty} \dd m' \,
	\frac{\partial \ucond}{\partial T}(T, \{ \mu \}; m', m')
	\\[0.25em]
	&\phantom{=\;}
	+
	\int_{\masss}^{\infty} \dd m' \,
	\frac{\partial \scond}{\partial T}(T, \{ \mu \}; m', m') \, .
\end{aligned}
\eeq

We would like to emphasize that Eqs.~\eqref{eq:entropy_density_final} and
\eqref{eq:entropy_density_final_2+1} are obtained without any approximation
and are therefore exact. Only the quark condensate as a function of
the quark mass (at fixed $T$ and $\{ \mu \}$) is needed to compute the
entropy density. This renders Eqs.~\eqref{eq:entropy_density_final} and
\eqref{eq:entropy_density_final_2+1} quite general and not constraint
to certain approaches---they are applicable as soon as the quark condensate
is available. This is particularly useful within the framework of DSEs, where
accessing the thermodynamic potential is extremely difficult and limited to
simple models with rainbow-ladder-like truncations
\cite{Blaschke:1997bj,Xu:2015jwa,Gao:2015kea}. The method presented here
allows us to bypass these limitations in order to compute thermodynamic
quantities regardless of the chosen truncation.%
\footnote{The method itself is truncation-independent but the obtained
results naturally not since the quantitative behavior of the quark
condensate, especially as a function of the quark mass, is truncation
dependent.}

\section{\label{sec:njl}NJL-model study}

To show that the method described in the previous section works effectively,
we use a two-flavor NJL model \cite{Nambu:1961tp,Nambu:1961fr} in mean-field
approximation, where the thermodynamic potential can be computed analytically.

The two-flavor NJL Lagrangian in mean-field approximation is given by
\cite{Klevansky:1992qe,Hatsuda:1994pi,Buballa:2003qv}
\beq
	\label{eq:njl_lagrangian}
	\calL_{\text{NJL}}^{\text{(mf)}}
	=
	\bar{\psi} \+ (\ii \+ \slashed{\partial} - M) \+ \psi
	-
	\frac{(M - m)^{2}}{4 \+ G} \, ,
\eeq
where $m$ is the current-quark mass, $M$ the constituent-quark mass, and $G$
denotes the coupling constant. The thermodynamic potential is thus simply the
noninteracting one \cite{Kapusta:2006pm} shifted by a field-independent term,
i.e.,
\beq
\begin{aligned}
	\label{eq:njl_omega}
	\!\! \OmegaNJL
	&=
	\frac{(M - m)^{2}}{4 \+ G}
	-
	2 \+ \Nf \+ \Nc \+
	\biggl[ \,
		\int^{\Lambda} \frac{\dd^{3} \vec{k}}{(2 \pi)^{3}} \, E_{\vec{k}}
		\\[0.25em]
		&\phantom{=\;} +
		T \int \frac{\dd^{3} \vec{k}}{(2 \pi)^{3}} \sum_{z = \pm 1}
		\log \Bigl( 1 + \ee^{-(E_{\vec{k}} + z \mu) / \+ T} \Bigr)
	\biggr] \+ ,
\end{aligned}
\eeq
with $E_{\vec{k}} = \sqrt{\vec{k}^{2} + M^{2}}$; $\Nf = 2$ and $\Nc = 3$ denote
the number of flavors and colors, respectively. We regularize the divergent
vacuum integral with a sharp three-momentum cutoff,
$\abs{\vec{k}} \leq \Lambda$, but leave the convergent medium contribution
unaltered. The physical constituent-quark mass is obtained by minimizing the
potential,
\beq
	\label{eq:njl_omega_minimize}
	\frac{\partial \+ \OmegaNJL}{\partial M}
	\stackrel{!}{=}
	0 \, ,
\eeq
and the quark condensate reads
\beq
	\label{eq:njl_condensate}
	\qcond_{\text{NJL}}^{\text{(mf)}}
	=
	-\frac{M - m}{2 \+ G} \, .
\eeq

Finally, the model is complete once the parameters are fixed. We use
$m = \SI{5.6}{\MeV}$, $\Lambda = \SI{587.9}{\MeV}$, and
$G = 2.44 \+ / \Lambda^{2}$. These values were determined in
Ref.~\cite{Buballa:2003qv} to yield a pion mass and decay constant of
$m_{\pi} = \SI{135}{\MeV}$ and $f_{\pi} = \SI{92.4}{\MeV}$ in vacuum.
The resulting constituent-quark mass in vacuum is
$M_{\text{vac}} = \SI{400}{\MeV}$.

\begin{figure}[t]
	\centering%
	\includegraphics{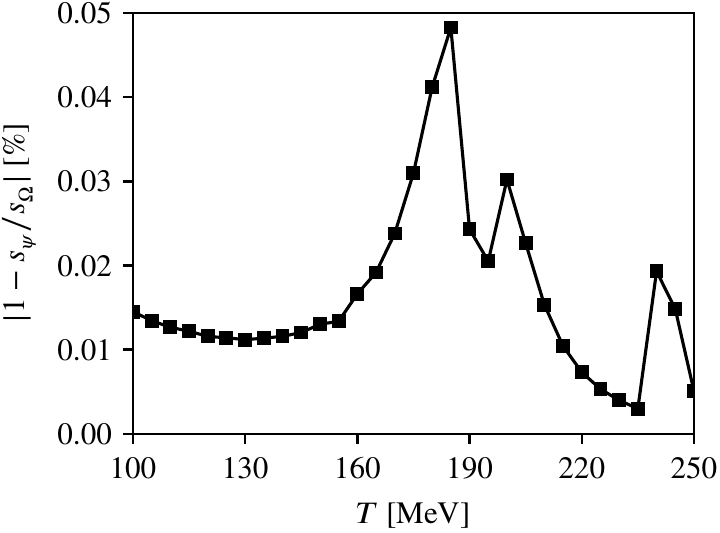}%
	\vspace{-1mm}%
	\caption{\label{fig:s_njl_comparison}%
		Relative error between the entropy density in a two-flavor
		NJL model obtained from the quark condensate ($s_{\psi}$) as
		described in Sec.~\ref{sec:method} and directly from the
		thermodynamic potential ($s_{\Omega}$).
	}
\end{figure}

We are now in a position to compute the entropy density first directly from the
thermodynamic potential \eqref{eq:njl_omega} using Eqs.~\eqref{eq:p_s_n}
and second via the method described in Sec.~\ref{sec:method}, i.e., by means of
Eq.~\eqref{eq:entropy_density_final}. As mentioned earlier, $\SYM = 0$ since
gluons are no active degrees of freedom in the NJL model. We find that both
results cannot be distinguished by the eye and show the relative error between
them in Fig.~\ref{fig:s_njl_comparison}. It is smaller than
$\SI{0.05}{\percent}$ across the whole covered temperature range. Thus, we are
confident that our method to obtain the entropy density from the quark
condensate is able to yield reliable results in the Dyson-Schwinger approach,
too.

\section{\label{sec:dse}Dyson-Schwinger equations}

\begin{figure}[t]
	\centering%
	\includegraphics[scale=1.1]{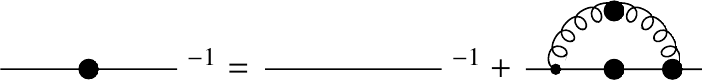}%
	\caption{\label{fig:qDSE}%
		The DSE for the quark propagator. Large filled circles denote dressed
		quantities; solid and wiggly lines represent quarks and gluons,
		respectively. There is a separate DSE for each quark flavor.
	}
\end{figure}

\begin{figure}[t]
	\centering%
	\includegraphics[scale=1.05]{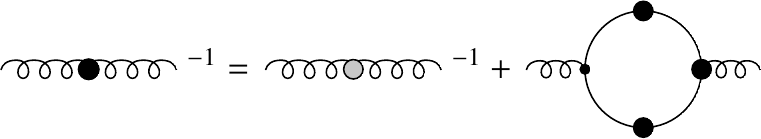}%
	\caption{\label{fig:gDSE}%
		Truncated gluon DSE. The gray circle denotes the quenched gluon
		propagator from the lattice, and the quark-loop diagram contains an
		implicit flavor sum. We use $\Nf = 2 + 1$ quark flavors in this work.
	}
\end{figure}

In the following, we briefly summarize our functional framework of
Dyson-Schwinger equations used to determine the dressed quark propagator.
From it, the quark condensate and eventually the entropy density are obtained.
We use the same setup as in our previous work \cite{Isserstedt:2019pgx} and
solve a coupled set of Landau-gauge Dyson-Schwinger equations, where the back
reaction of the quarks onto the Yang-Mills sector is explicitly taken into
account.

The dressed quark propagator $S_{f}$ for a flavor $f$ at nonzero temperature
$T$ and quark chemical potential $\mu_{f}$ is the solution of the DSE%
\footnote{We work in four-dimensional Euclidean space-time with Hermitian gamma
matrices obeying $\{ \gamma_\mu, \gamma_\nu \} = 2 \+ \delta_{\mu\nu}$ and use
the Matsubara formalism to describe the system at nonzero temperature.}
\beq
	\label{eq:dse_quark}
	S_{f}^{-1}(q)
	=
	Z_{2}^{f} \bigl( \+ \ii \+ \wtilde{\omega}_{q} \gamma_4
	+
	\ii \+ \vec{\gamma} \cdot \vec{q} 
	+
	Z_{m}^{f} m_{f} \bigr)
	+
	\Sigma_{f}(q) \, ,
\eeq
which is depicted in Fig.~\ref{fig:qDSE}. Here,
$q = (\wtilde{\omega}_{q}, \vec{q})$ is the four-momentum with
$\wtilde{\omega}_{q} = \omega_{q} + \ii \+ \mu_f$ and fermionic
Matsubara frequencies $\omega_{q} = (2 \+ \ell_{q} + 1) \+ \pi T$,
$\ell_{q} \in \bbZ$. Furthermore, $Z_{2}^f$ and $Z_{m}^{f}$ denote the
wave function and mass renormalization constants; $m_f$ is the
renormalized current-quark mass. The inverse dressed quark propagator
is given by
\beq
	\label{eq:quark_propagator}
	S_{f}^{-1}(q) 
	=
	\ii \+ \wtilde{\omega}_{q} \gamma_{4} \+ C_{f}(q)
	+
	\ii \+ \vec{\gamma} \cdot \vec{q} \+ A_{f}(q)
	+
	B_{f}(q) \, ,
\eeq
with scalar dressing functions $C_{f}$, $A_{f}$, and $B_{f}$. They carry the
nonperturbative information, thus having a nontrivial momentum dependence, and
depend on temperature and chemical potential as well. A further dressing
function corresponding to the tensor structure
$\gamma_4 \+ \vec{\gamma} \cdot \vec{q}$ is in principle possible but
qualitatively negligible \cite{Contant:2017gtz}. The explicit form of the
quark self-energy appearing in Eq.~\eqref{eq:dse_quark} reads
\beq
	\label{eq:quark_self_energy}
	\Sigma_f(q)
	=
	g^{2} \+
	\frac{4}{3}
	\frac{Z_{2}^{f}}{\wtilde{Z}_{3}}
	\sum_{\ell_{k}}
	\int \frac{\dd^{3} \vec{k}}{(2 \pi)^{3}} \,
	\gamma_{\nu} \+
	D_{\nu\sigma}(k - q) \+
	S_{f}(k) \+
	\Gamma_{\sigma}^{f}(k, q) \, ,
\eeq
with the dressed gluon propagator $D_{\mu\nu}$, dressed quark-gluon vertex
$\Gamma_{\sigma}^{f}$, strong coupling constant $g$, ghost renormalization
constant $\wtilde{Z}_{3}$, and loop momentum
$k = (\wtilde{\omega}_{k}, \vec{k})$.
The prefactor $4/3$ stems from the color trace.

In order to solve the quark DSE self-consistently for the dressed quark
propagator, we need to specify the dressed gluon propagator and quark-gluon
vertex. The truncation used in this work evolved gradually
(see Ref.~\cite{Fischer:2018sdj} and references therein)
and is characterized as follows. First, we use temperature-dependent lattice
data for the quenched gluon propagator \cite{Fischer:2010fx,Maas:2011ez} as
input and incorporate unquenching effects by explicitly evaluating the
quark-loop diagram for each of the $\Nf = 2 + 1$ quark flavors considered here.
This results in the DSE for the unquenched gluon propagator as shown in
Fig.~\ref{fig:gDSE}. Consequently, the quark and gluon DSEs are nontrivially
coupled and need to be solved simultaneously. This construction allows for a
dependence of the gluon on temperature and chemical potential controlled by
QCD dynamics rather than modeling. Second, we use an ansatz for the dressed
quark-gluon vertex motivated by its known perturbative running in the
ultraviolet combined with an approximate form of the Slavnov-Taylor identity
in the infrared based on the Ball-Chiu vertex construction \cite{Ball:1980ay}.
Since our setup is identical to the one used recently, we shall not repeat
explicit expressions regarding the truncation for the sake of brevity and
refer the reader to Ref.~\cite{Isserstedt:2019pgx}.

The free parameters of the truncation are the infrared strength of the vertex
ansatz and the quark masses. They are fixed to yield a pseudocritical chiral
transition temperature at vanishing chemical potential of
$\Tc = \SI{156}{\MeV}$, defined by the inflection point of the light quark
condensate with temperature, in agreement with lattice results
\cite{Borsanyi:2010bp,Bazavov:2011nk,Bellwied:2015rza,
Bonati:2018nut,Bazavov:2018mes}.
We work in the isospin-symmetric limit $\massu = \massd$,
$\muu = \mud \equiv \mu$ and choose $\mus = 0$ for simplicity. This implies
that the baryon chemical potential is given by $\muB = 3 \+ \mu$.

Finally, after solving the coupled set of quark and gluon DSEs for the 
propagators, the corresponding quark condensate is obtained via
\beq
	\label{eq:condensate}
	\qcond_{f}
	=
	-3 \+ Z_{2}^{f} Z_{m}^{f} \sum_{\ell_{k}} \int
	\frac{\dd^{3} \vec{k}}{(2 \pi)^{3}} \Tr\bigl[ S_f(k) \bigr] \+ .
\eeq
Its gauge invariance and hence gauge invariance of the results is
guaranteed by the Landau-Khalatnikov-Fradkin transformations
\cite{Landau:1955zz,Fradkin:1955jr}.

\section{\label{sec:results}Results and discussion}

\begin{figure}[t]
	\centering%
	\includegraphics{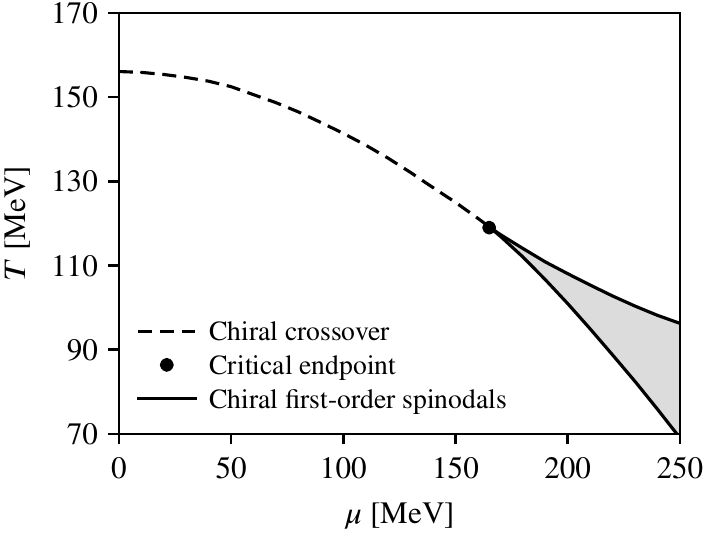}%
	\vspace{-1mm}%
	\caption{\label{fig:phase_diagram}%
		Phase diagram of QCD obtained from the DSE setup used in this work.
	}
\end{figure}

For the sake of completeness, we first recapitulate the phase diagram obtained
with the DSE setup described in Sec.~\ref{sec:dse}. It is shown in
Fig.~\ref{fig:phase_diagram}, and we refer the reader to
Ref.~\cite{Isserstedt:2019pgx} for more details. The chiral crossover line
(dashed black) starting at $\Tc(\mu = 0) = \SI{156(1)}{\MeV}$ becomes steeper
with increasing chemical potential and terminates in a second-order critical
endpoint (CEP) (solid black circle) located at
$(\TCEP, \muCEP) = (119 \pm 2, 165 \pm 2) \, \si{\MeV}$.
This corresponds to a ratio of
$(\muB / \+ T)_{\text{CEP}} \approx 4.2$.
The errors are purely numerical. Beyond the CEP, we find the coexistence region
of a first-order transition (shaded area) bounded by spinodals (solid black).

We now focus on our thermodynamic results obtained with the method described
in Sec.~\ref{sec:method} used with condensate data computed from the
($2 + 1$)-flavor DSE framework summarized in the previous section.
The starting point is the entropy density obtained via
Eq.~\eqref{eq:entropy_density_final_2+1}. Lattice simulations of pure
Yang-Mills theory show that there is basically no sizable contribution to
$\SYM / \+ T^3$ for $T \lesssim \Tc^{\text{YM}} \approx \SI{270}{\MeV}$
\cite{Boyd:1996bx,Borsanyi:2012ve}, i.e., for temperatures around and below
the critical Yang-Mills temperature. Thus, to a good approximation, we set
$\SYM = 0$ in Eq.~\eqref{eq:entropy_density_final_2+1} for the temperature
range covered in this work.

\subsection{Zero chemical potential}

\begin{figure}[t]
	\centering%
	\includegraphics{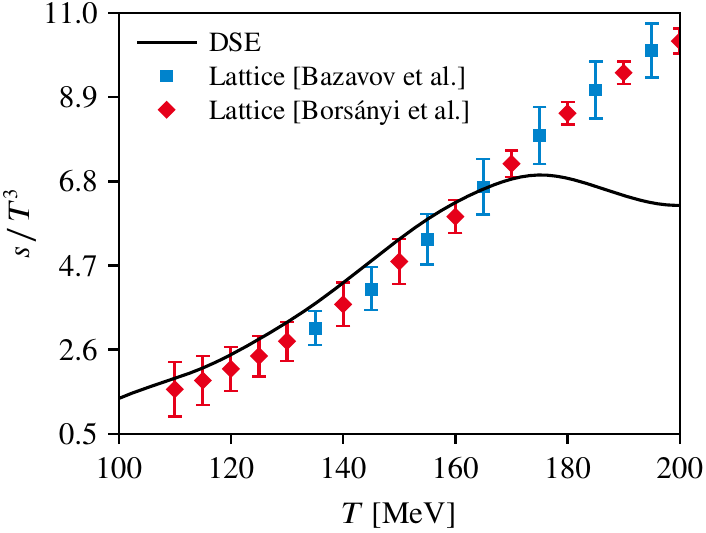}%
	\\[2mm]%
	\includegraphics{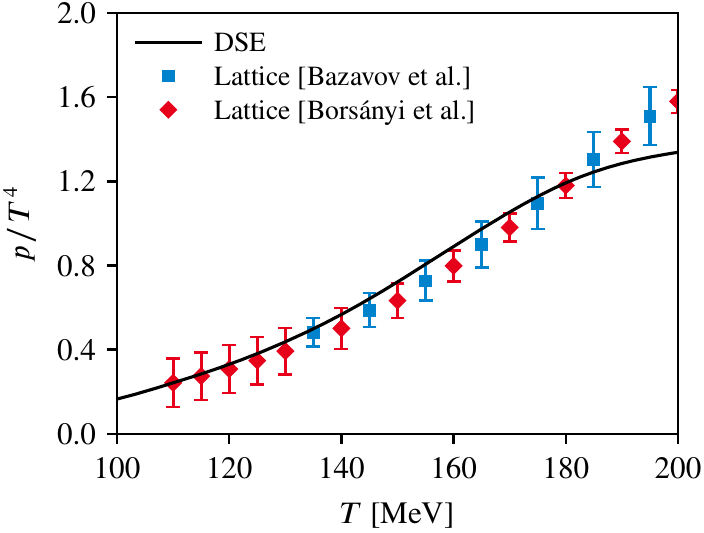}%
	\\[2mm]%
	\includegraphics{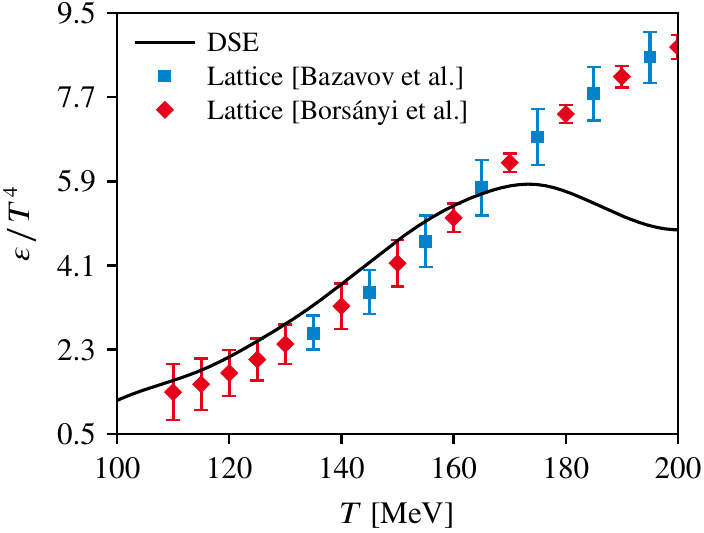}%
	\vspace{-1mm}%
	\caption{\label{fig:spe}%
		Entropy density (top), pressure (center), and energy density (bottom)
		at vanishing chemical potential. The lattice data is taken from
		Refs.~\cite{Borsanyi:2013bia,Bazavov:2014pvz}.
	}
\end{figure}

Our result for the entropy density (scaled to $T^3$) at vanishing chemical
potential is shown in the upper diagram of Fig.~\ref{fig:spe} (solid black line)
compared to results from lattice QCD \cite{Borsanyi:2013bia,Bazavov:2014pvz}
(colored symbols). Up to $T \approx \SI{175}{\MeV}$, it is a monotonically
increasing function of temperature, and the agreement with lattice data is
satisfying. Beyond that temperature, the entropy density starts to decrease.
This unphysical behavior can be traced back to a deficiency in our vertex
ansatz and became apparent already in the calculation of quark and baryon
number fluctuations \cite{Isserstedt:2019pgx}. At the moment, we take only the
leading Dirac structure $\gamma_\sigma$ of the dressed quark-gluon vertex into
account. However, the full vertex $\Gamma_{\sigma}^f$ contains twenty-four
different Dirac tensor structures (in Landau gauge) with half of them reacting
strongly to the (partial) restoration of chiral symmetry around and above the
pseudocritical chiral transition temperature. These terms are missing in our
current setup. Their inclusion would cause a continuous weakening of the
quark-gluon interaction for $T \gtrsim \Tc$, thereby resolving the issue of a
decreasing entropy density at high temperatures.%
\footnote{This effect could be mimicked by making the interaction strength of
our vertex ansatz temperature dependent. An analogous modification is needed in
rainbow-ladder models as well to achieve proper results for thermodynamics
above $\Tc$ \cite{Gao:2015kea}.}
\begin{figure*}[t]
	\centering%
	\includegraphics{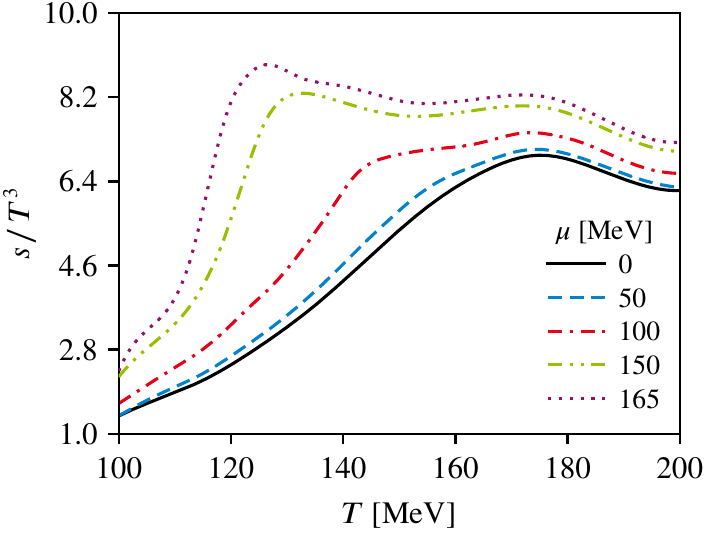}%
	\hspace{2mm}%
	\includegraphics{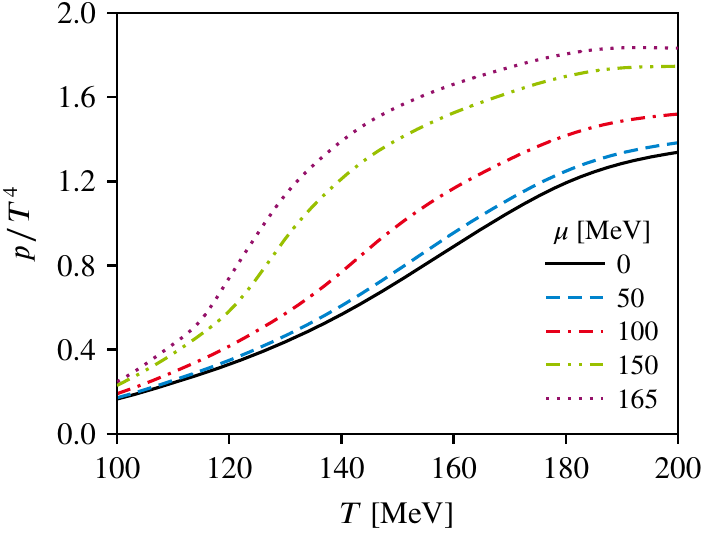}%
	\\[2mm]%
	\includegraphics{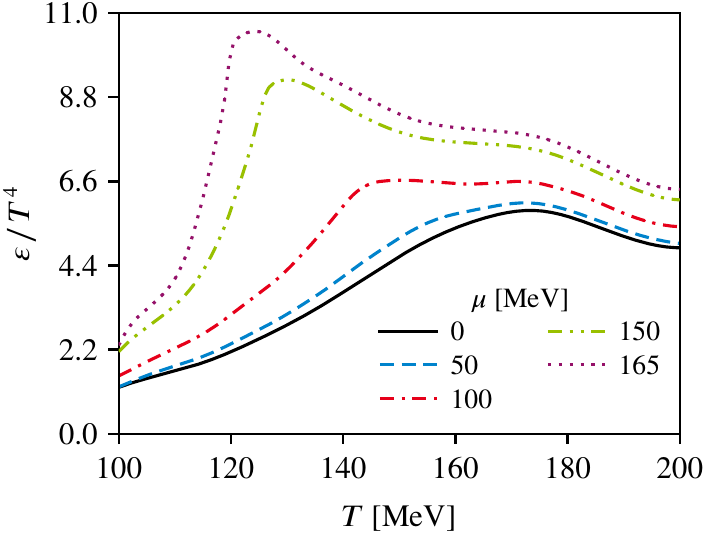}%
	\hspace{2mm}%
	\includegraphics{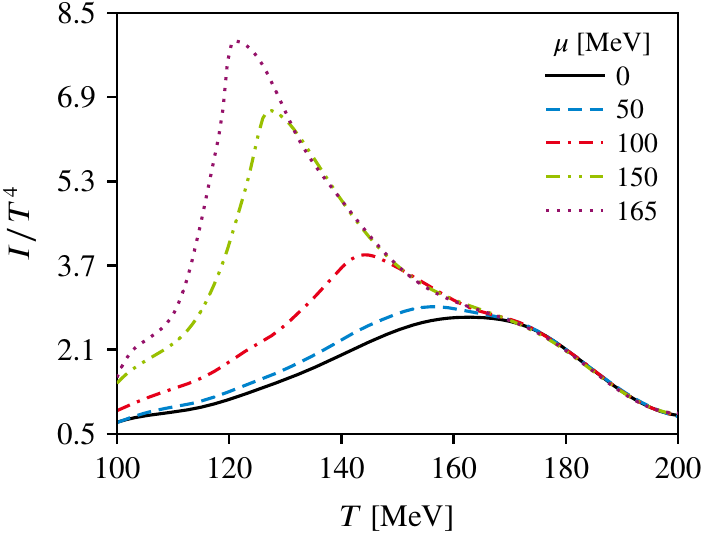}%
	\vspace{-1mm}%
	\caption{\label{fig:spe_mu}%
		Entropy density (top left), pressure (top right), energy density
		(bottom left), and interaction measure (bottom right) as functions 
		of temperature for different chemical potentials up to the CEP.
	}
\end{figure*}
A proper treatment of the full vertex goes beyond the scope of this work
but we note, however, that our setup yields satisfying results in the
temperature range $100$--$\SI{160}{\MeV}$ below and around $\Tc$.

From the entropy density we obtain the pressure via integration, see
Eq.~\eqref{eq:pressure_zero_mu}, and use
$p(T_{0}) / \+ T_{0}^{4} = 0.242$ at $T_{0} = \SI{110}{\MeV}$
\cite{Borsanyi:2013bia}. The result depicted in the middle diagram of
Fig.~\ref{fig:spe} is in good agreement with the lattice but starts to deviate
for $T \gtrsim \SI{185}{\MeV}$. This is inherited from the erroneous
high-temperature behavior of the entropy density. The pressure saturates at
$p / p_{\text{SB}} \approx 0.3$, where
\beq
	\label{eq:p_SB}
	\frac{p_{\text{SB}}}{T^{4}}
	=
	\frac{19 \pi^{2}}{36}
	+
	\left( \frac{\mu}{T} \right)^{2}
	+
	\frac{1}{2 \pi^{2}} \left( \frac{\mu}{T} \right)^{4}
\eeq
is the Stefan-Boltzmann pressure of an ideal gas of massless quarks and gluons.
Finally, combining the entropy density and pressure we obtain the energy
density, Eq.~\eqref{eq:epsilon}, shown in the lower diagram of
Fig.~\ref{fig:spe}. Since it is a combination of $s$ and $p$, the agreement
with lattice results is reasonable for temperatures below and around $\Tc$
while a decreasing behavior stemming from the entropy density is found at
high temperatures.

Ignoring the high-temperature artifacts, one can also define
the pseudocritical chiral transition temperature as the inflection point of,
e.g., the pressure with temperature. We find $\Tc^{\+(p)} = \SI{157}{\MeV}$
consistent with \SI{156}{\MeV} obtained from the light quark condensate.

\subsection{Nonzero chemical potential}

We now turn to nonvanishing chemical potential and show our results in
Fig.~\ref{fig:spe_mu}. The entropy density is depicted in the upper left diagram
as a function of temperature for different chemical potentials starting from
zero up to the critical endpoint value $\muCEP = \SI{165}{\MeV}$. A bulge
develops around the pseudocritical chiral transition temperature and becomes
more pronounced with increasing chemical potential. Close to and across the CEP,
we find a strong increase of the entropy density with temperature, and the slope
becomes maximal at $T = \TCEP$. The incorrect high-temperature behavior
persists and becomes nonmonotonic.

The pressure follows again via integration according to
Eq.~\eqref{eq:pressure_nonzero_mu} while the number density is computed as
described in Ref.~\cite{Isserstedt:2019pgx}. As seen in the upper right
diagram, $p$ gets larger with increasing chemical potential across the whole
temperature range, but the changes are less noticeable at low temperatures.
For chemical potentials close to the CEP, a kink starts to form at the
corresponding transition temperature $\Tc(\mu)$. After that, the pressure
rises stronger with a steeper slope as a function of $T$; most pronounced and
noticeable directly at the CEP. The pressure is, however, a smooth function of
temperature for all $\mu$ up to $\muCEP$. These results are consistent with FRG
results from the (Polyakov-loop enhanced) quark-meson model; see, e.g.,
Ref.~\cite{Herbst:2013ail}.

The effect of a nonzero and increasing chemical potential is most prominent
in the energy density (bottom left diagram) due to the additional
number-density term $\mu \+ n(T, \mu)$ from the Legendre transform; see
Eq.~\eqref{eq:epsilon}. Its steep rise close to CEP indicates a rapid increase
of degrees of freedom from hadrons to quarks and gluons. This behavior carries
over to the interaction measure (bottom right diagram), which reacts strongly to
chemical potential, too. It is shape consistent with lattice results at small
chemical potential and experiences a strong increase from intermediate chemical
potentials onwards to the CEP. There, at $\mu = \muCEP$, the slope becomes
infinite at the corresponding critical temperature $\TCEP$. The peaklike
structure of $I / \, T^4$ close to and at the CEP with a large magnitude
indicates that nonperturbative effects are manifest in this region of the
phase diagram.

\subsection{The first-order phase boundary}

With the pressure as a function of temperature and chemical potential at hand,
the next obvious step is in principle the determination of the first-order
phase boundary which lies between the spinodals (shaded area in
Fig.~\ref{fig:phase_diagram}). Unfortunately, as discussed in the following,
there we hit a limitation of the method described in Sec.~\ref{sec:method}.

In order to locate the first-order phase boundary, one considers the
pressure difference
\beq
	\label{eq:pressure_difference}
	\calB(T, \mu) = \pN(T, \mu) - \pW(T, \mu)
\eeq
between the chirally broken Nambu (N) and the chirally symmetric Wigner (W)
phase for a fixed $T < \TCEP$ as a function of the chemical potential.
Clearly, $\calB(T, \mu)$ is only defined up to the chemical potential
$\mucN = \mucN(T)$, above which the Nambu solution does not exist anymore and
only the Wigner solution can be found. The physically realized phase maximizes
the pressure: $\calB(T, \mu) > 0$ indicates that the Nambu phase is more stable
than the Wigner phase and vice versa. Therefore, $\calB(T, \mu) = 0$ defines
the phase boundary. By finding the root of $\calB(T, \mu)$ with respect to
$\mu$ for various $T \in [0, \TCEP)$, one can draw the first-order phase
boundary in the phase diagram.

In our approach, the pressure difference is explicitly given by
\beq
	\label{eq:pressure_difference_explicitly}
	\calB(T, \mu)
	=
	\Deltamuz(T) - \Deltamunz(T, \mu) \, ,
\eeq
with the functions
\beq
\begin{aligned}
	\label{eq:pressure_difference_part1}
	\Deltamuz(T)
	&=
	\pN(T_0, 0) - \pW(T_0, 0)
	\\[0.5em]
	&\phantom{=\;}+
	\int_{T_0}^{T} \dd T\+' \, \bigl[ \sN(T\+', 0) - \sW(T\+', 0) \bigr]
\end{aligned}
\eeq
and
\beq
	\label{eq:pressure_difference_part2}
	\Deltamunz(T, \mu)
	=
	\int_{0}^{\mu} \dd \mu' \, \bigl[ \nW(T, \mu') - \nN(T, \mu') \bigr] \+ .
\eeq
The first part \eqref{eq:pressure_difference_part1} is evaluated at
$\mu = 0$ only; i.e., it does not depend on chemical potential. Moreover,
$\Deltamuz(T)$ is positive for all $T \in [0, \TCEP)$ since this region of the
phase diagram is well within the hadronic phase where the Nambu solution is
realized. Next, we find both in the NJL model as well as in our DSE setup
that the number density as a function of $\mu$ (at fixed $T < \TCEP$) of the
Wigner phase is generally larger than the number density of the Nambu phase:
$\nW(T, \mu) - \nN(T, \mu) > 0$, and therefore, $\Deltamunz(T, \mu) > 0$ for
all $\mu \in [0, \mucN]$. Furthermore, $\Deltamunz(T, \mu)$ is a monotonically 
increasing function of the chemical potential.
It follows that there exists $\mucfirst \in [0, \mucN]$---the location of the
first-order phase boundary---where $\Deltamunz(T, \mucfirst) = \Deltamuz(T)$,
and consequently $\calB(T, \mucfirst) = 0$.

From the above discussion, it becomes apparent that the correct location of
the phase boundary depends crucially on the value of $\Deltamuz$. In
particular, we need the pressure difference $\pN(T_0, 0) - \pW(T_0, 0)$ as
input while all other quantities in Eqs.~\eqref{eq:pressure_difference_part1}
and \eqref{eq:pressure_difference_part2} are computed from DSEs. Alas, lattice
QCD provides to our knowledge only the physical Nambu pressure $\pN(T_0, 0)$
but not the unphysical Wigner pressure $\pW(T_0, 0)$ at a reference
temperature $T_0$. Thus, we are not able to obtain a reliable location of the
phase boundary in the first-order region of the phase diagram. However, we
would like to note that this is not an inherent flaw of our method but more
that we hinge on the availability of an external input parameter.

\section{\label{sec:summary}Summary and conclusions}

In this work, we have studied the thermodynamics of strong-interaction matter
within the DSE framework. We proposed a method to compute the entropy density
solely from the quark condensate; a subsequent integration yields the pressure.
The key feature of the method is that no approximation is used during its
derivation and only the quark condensate is needed as input. This is
particularly useful for DSEs, where accessing the thermodynamic potential is
extremely difficult and limited to truncations of the rainbow-ladder type. Even
then, the proper removal of the quartic divergence contained in the potential
is a nontrivial task. The proposed method provides a truncation-independent and
straightforward way to compute thermodynamic quantities as soon as one gets hold
of the quark condensate. The results, however, are truncation dependent since
the quantitative behavior of the quark condensate depends on the chosen
truncation.

That the method works effectively and yields reliable results was shown
successfully using a NJL model. Then, we used condensate data obtained from a
coupled set of Dyson-Schwinger equations for the quark and gluon propagators of
($2 + 1$)-flavor QCD within a truncation scheme used and discussed previously
\cite{Isserstedt:2019pgx,Fischer:2018sdj} to obtain the pressure, entropy
density, energy density, and interaction measure from zero chemical potential
up to the CEP as functions of temperature. These thermodynamic results are, to
our knowledge, the first ones obtained from DSEs with a beyond-rainbow-ladder
truncation---emphasizing the usefulness of the method presented in
Sec.~\ref{sec:method}.

At vanishing chemical potential, we find that our results for the pressure,
entropy density, and energy density are in very good agreement with lattice
QCD for temperatures below and around the pseudocritical chiral transition
temperature. However, at high temperatures, we observe an unphysical decrease
of the entropy density with temperature. This erroneous behavior is rooted in
the vertex ansatz of our DSE setup. Results at nonzero chemical potential show
the expected behavior if one increases the chemical potential and are in
qualitative agreement with results obtained within the FRG applied to the
(Polyakov-loop enhanced) quark-meson model.

Unfortunately, we are not able to determine the first-order phase boundary
below the CEP. For that, we need the pressure difference between the Nambu and
Wigner phase at a reference temperature and at vanishing chemical potential
as input. However, this quantity is as far as we know not provided by
lattice-QCD calculations.

Finally, the results obtained in this work together with the ones on
quark and baryon number fluctuations \cite{Isserstedt:2019pgx} made clear
that an elaborate dressed quark-gluon vertex is needed for proper thermodynamics
at high temperatures and/or densities. Especially terms in the vertex that
react strongly to the (partial) restoration of chiral symmetry at high
temperatures and/or chemical potential are of crucial importance. This
extension of our current setup could be guided, e.g., by vacuum results for
the dressed quark-gluon vertex or by an explicit calculation of (parts of) the
vertex in medium \cite{Williams:2014iea,Contant:2018zpi}. A different yet
complementary approach, which takes more vertex structures into account, is the
FRG-assisted difference-DSE method proposed in
Refs.~\cite{Gao:2020qsj,Gao:2020fbl}, where nonzero temperature and chemical
potential is treated as a fluctuation around the vacuum.


\begin{acknowledgments}
	We are grateful to Pascal J.~Gunkel, Bernd-Jochen Schaefer, and Richard
	Williams for fruitful discussions and thank Christian A.~Welzbacher for
	contributions at an early stage of this work. Furthermore, we thank Pascal
	J.~Gunkel for a thorough reading of the manuscript. This work was supported
	by the Helmholtz Graduate School for Hadron and Ion Research for FAIR, the
	GSI Helmholtzzentrum f\"{u}r Schwerionenforschung, and the BMBF under
	Contract No.~05P18RGFCA. We acknowledge computational resources provided
	by the HPC Core Facility and the HRZ of the Justus-Liebig-Universit\"{a}t
	Gie\ss{}en. Feynman diagrams were drawn with \textit{JaxoDraw}
	\cite{Binosi:2008ig}.
\end{acknowledgments}


\urlstyle{same}
\bibliography{ThermodynamicsCondensateBibliography}

\end{document}